\documentclass[10pt,conference]{IEEEtran}
\usepackage{cite}
\usepackage{amsmath,amssymb,amsfonts}
\usepackage{algorithmic}
\usepackage{graphicx}
\usepackage{textcomp}
\usepackage{xcolor}
\usepackage[hyphens]{url}
\usepackage{booktabs}
\usepackage{pifont}
\usepackage{siunitx}
\usepackage{amsmath}
\usepackage{algorithm2e}
\usepackage{multirow}
\usepackage{fancyhdr}
\usepackage{float}
\usepackage{setspace}
\usepackage{listings}
\usepackage{balance}
\usepackage{wrapfig}
\usepackage{marginnote}
\usepackage{caption}
\usepackage{floatflt}
\definecolor{gfored}{rgb}{0.580, 0.050, 0.211}
\definecolor{ao}{rgb}{0.007, 0.520, 0.867}
\definecolor{yt}{rgb}{0.875, 0.568, 1.000}
\definecolor{moegi}{rgb}{0.357, 0.537, 0.188}
\definecolor{jl}{rgb}{1.0, 0.2, 0.8}
\definecolor{brown(web)}{rgb}{0.65, 0.16, 0.16}
\definecolor{bisque}{rgb}{1.0, 0.89, 0.77}

\newif\ifsqueezefigs
\squeezefigstrue

\ifsqueezefigs
\makeatletter
\g@addto@macro{\normalsize}{%
  \setlength{\abovedisplayskip}{2pt plus 1pt minus 1pt}
  \setlength{\belowdisplayskip}{2pt plus 1pt minus 1pt}
  \setlength{\abovedisplayshortskip}{0pt}
  \setlength{\belowdisplayshortskip}{0pt}
  \setlength{\intextsep}{2pt plus 1pt minus 1pt}
  \setlength{\textfloatsep}{3pt plus 1pt minus 1pt}
  \setlength{\dbltextfloatsep}{3pt plus 1pt minus 1pt}
  \setlength{\skip\footins}{4pt plus 1pt minus 1pt}}
\makeatother
\fi

\newif\ifdraft
\draftfalse

\ifdraft
    \usepackage[colorinlistoftodos,prependcaption,textsize=small]{todonotes}
    \paperwidth=\dimexpr \paperwidth + 4cm\relax
    \oddsidemargin=\dimexpr\oddsidemargin + 2cm\relax
    \evensidemargin=\dimexpr\evensidemargin + 2cm\relax
    \marginparwidth=\dimexpr \marginparwidth + 2cm\relax

    \newcommand{\ominline}[1]{\textcolor{red}{\textbf{[@om: }#1\textbf{]}}}
    \newcommand{\ombox}[1]{\todo[size=\scriptsize, linecolor=red, bordercolor=red, backgroundcolor=white]{\textcolor{red}{\textbf{@om:} #1}}}

    \newcommand{\agycomment}[1]{\todo[size=\scriptsize, linecolor=orange, bordercolor=orange, backgroundcolor=white]{\textcolor{gfored}{\textbf{@gy:} #1}}}
    \newcommand{\agyinline}[1]{\textcolor{gfored}{\textbf{[@agy: }#1\textbf{]}}}

    \newcommand{\atbcomment}[1]{\todo[size=\scriptsize, linecolor=orange, bordercolor=orange, backgroundcolor=white]{\textcolor{ao}{\textbf{@atb:} #1}}}

    \newcommand{\hluoinline}[1]{\textcolor{moegi}{\textbf{[@hluo: }#1\textbf{]}}}
    \newcommand{\hluobox}[1]{\todo[size=\scriptsize, linecolor=orange, bordercolor=orange, backgroundcolor=white]{\textcolor{moegi}{\textbf{@hluo:} #1}}}

    \newcommand{\yctcomment}[1]{\todo[size=\scriptsize, linecolor=orange, bordercolor=orange, backgroundcolor=white]{\textcolor{yt}{\textbf{@yct:} #1}}}

    \newcommand{\joel}[1]{\textcolor{jl}{#1}}
    \newcommand{\joelcomment}[1]{\todo[size=\scriptsize,linecolor=orange,bordercolor=orange,backgroundcolor=white]{\textcolor{jl}{\textbf{@joel:} #1}}}

\else

    \newcommand{\ominline}[1]{}
    \newcommand{\ombox}[1]{}

    \newcommand{\agycomment}[1]{}
    \newcommand{\agyinline}[1]{}

    \newcommand{\atbcomment}[1]{}

    \newcommand{\hluoinline}[1]{}
    \newcommand{\hluobox}[1]{}

    \newcommand{\yctcomment}[1]{}

    \newcommand{\joel}[1]{{#1}}
    \newcommand{\joelcomment}[1]{}

\fi

\newif\ifrebuttal
\rebuttaltrue

\ifrebuttal
\usepackage[colorinlistoftodos,prependcaption,textsize=small]{todonotes}

\definecolor{darkred}{rgb}{0.9, 0.0, 0.0}

\definecolor{darkblue}{rgb}{0.0, 0.0, 0.85}

\fi

\newcommand*\DRAMCMD[1]{\texttt{#1}}
\newcommand*\DRAMTIMING[1]{t\textsubscript{#1}}

\newcounter{obs}
\newcounter{tkw}
\usepackage{glossaries}

\newacronym{vdd}{$V_{DD}$}{supply voltage}
\newacronym{vpp}{$V_{PP}$}{wordline voltage}
\newacronym{vwl}{$V_{PP}$}{wordline voltage}
\newacronym{vgs}{$V_{GS}$}{gate-to-source voltage}
\newacronym{vth}{$V_{TH}$}{the voltage threshold that the bitline voltage should exceed for the activation to be reliably completed}
\newacronym{gnd}{$GND$}{ground}

\newacronym{ber}{$BER$}{the fraction of DRAM cells that experience bitflips in a DRAM row}
\newacronym{acmin}{$AC_{min}$}{the minimum number of total aggressor row activations to cause at least one bitflip}
\newacronym{ac}{$AC$}{activation count}
\newacronym{rblast}{$r_{Blast}$}{blast radius}
\newacronym{iqr}{$IQR$}{interquartile range}

\newacronym{trcd}{\DRAMTIMING{RCD}}{
{the minimum time between opening a row with an \DRAMCMD{ACT} command and accessing the row buffer}
}
\newacronym{trp}{\DRAMTIMING{RP}}{
{the minimum time between sending a \DRAMCMD{PRE} command and opening a row with an \DRAMCMD{ACT} command}
}
\newacronym{tras}{\DRAMTIMING{RAS}}{
{the minimum time between opening a row with an \DRAMCMD{ACT} command and closing the row with a \DRAMCMD{PRE} command}
}
\newacronym{trefi}{\DRAMTIMING{REFI}}{the \joel{default} time interval \joel{between consecutive \DRAMCMD{REF} commands}}
\newacronym{trefw}{\DRAMTIMING{REFW}}{the maximum time window between two consecutive refresh operations targeting {the same} row}

\def\BibTeX{{\rm B\kern-.05em{\sc i\kern-.025em b}\kern-.08em
    T\kern-.1667em\lower.7ex\hbox{E}\kern-.125emX}}

\author{
{Onur Mutlu}%
\vspace{-3pt}
\\
\emph{ETH Z{\"u}rich}%
\vspace{-25pt}
}

\title{\LARGE{\emph{Retrospective:} An Experimental Study of \\\vspace{-3pt} Data Retention Behavior in Modern DRAM Devices: \\\vspace{-3pt} Implications for Retention Time Profiling Mechanisms\vspace{-10pt}}}

\begin{document}
\maketitle
\thispagestyle{plain}
\pagestyle{plain}
\setstretch{0.8}

\begin{abstract}

DRAM is the prevalent main memory technology used in almost all computers. Data is represented as charge in a DRAM cell. Unfortunately, a DRAM cell loses its stored charge over time and thus needs to be refreshed periodically. How often a cell needs to be refreshed depends on its minimum data retention time, which is dependent on various factors.  Accurately identifying the minimum data retention time of each DRAM cell is necessary to 1) correctly determine the minimum refresh rate of a DRAM chip to maintain data integrity, and 2) enable techniques that eliminate unnecessary refresh operations by refreshing each DRAM row at the minimum refresh rate it needs for reliable operation.

Our ISCA 2013 paper~\cite{dram-isca2013} provides a fundamental empirical understanding of two major factors that make it very difficult to determine the minimum data retention time of a DRAM cell, based on the first comprehensive experimental characterization of retention time behavior of  a large number of modern commodity DRAM chips from 5 major vendors. We study the prevalence, effects, and technology scaling characteristics of two significant phenomena: 1) {\em data pattern dependence (DPD)}, where the minimum retention time of a DRAM cell is affected by data stored in other DRAM cells, and 2) {\em variable retention time (VRT)}, where the minimum retention time of a DRAM cell changes unpredictably over time. To this end, we built a flexible FPGA-based testing infrastructure to test DRAM chips, which has enabled a large amount of further experimental research in DRAM. Our ISCA 2013 paper's results using this infrastructure clearly demonstrate that DPD and VRT phenomena are significant issues that must be addressed for correct operation in DRAM-based systems and their effects are getting worse as DRAM scales to smaller technology node sizes. Our work also provides ideas on how to accurately identify data retention times in the presence of DPD and VRT, e.g., online profiling with error correcting codes, which later works examined and enabled. Most modern DRAM chips now incorporate ECC, especially to account for VRT effects~\cite{kang.memoryforum14}. 

This short retrospective provides a brief analysis of our ISCA 2013 paper and its impact. We describe why we did the work, what we found and its implications, what the findings as well as the infrastructure we built to discover them have enabled in later works, and our thoughts on what the future may bring. 

\end{abstract}

\section{Background} 
\vspace{-4pt}

My group has been working on the DRAM refresh problem since 2010 and our major work RAIDR~\cite{liu2012raidr} was published at ISCA 2012. Our goal in RAIDR was to eliminate unnecessary refresh operations at low cost by refreshing each DRAM row only as frequently as required by the minimum data retention time of the row. As described in a separate retrospective in this issue, our RAIDR work demonstrated large performance improvements and energy savings with a simple memory controller based implementation. However, we were not satisfied with the simplistic retention time profiling mechanism assumed in RAIDR (which was also assumed in other prior works). RAIDR relied on accurate identification of the minimum data retention time of every DRAM cell, which we thought was a difficult task. We wanted to make such retention time profiling practical. So, we set out to rigorously  understand the difficulty of DRAM data retention time identification using an empirical approach. No prior work at the time provided real data on the retention characteristics of state-of-the-art DRAM chips, let alone a detailed empirical analysis of major problems that make retention time profiling challenging and how DRAM technology scaling affects those challenges. In fact, no infrastructure to study these characteristics existed (or was available to us). We decided to build our own FPGA-based infrastructure to characterize real DRAM chips in a flexible manner so that we could change the refresh rate, data patterns, and other major parameters. This infrastructure, which took us more than a year to build and which we later open sourced as SoftMC~\cite{hassan2017softmc, softmc-safarigithub} and DRAM Bender~\cite{olgun2023drambender, safari-drambender}, enabled us (and others) to empirically study and understand many interesting characteristics of modern DRAM chips over the course of more than a decade. %

Our ISCA 2013 paper is a product of this goal and effort. Our work was generously supported especially by the Samsung DRAM Design Team and Intel Memory Architecture Labs, both technically and funding-wise. With close technical support from Intel (especially Chris Wilkerson, who is a co-author), we built our FPGA-based DDR3 DRAM testing infrastructure. Two of my students (also co-authors) and I spent part of the summer of 2012 at Intel to work closely with our collaborators. During this timeframe, we finalized the calibration and stabilization of our infrastructure. We performed many experiments to study both well-known properties of retention time characteristics (e.g., temperature dependence) as well as less well studied characteristics (e.g., DPD and VRT phenomena) of modern DRAM chips at a scale that was not reported before. We were especially interested in empirically understanding how technology scaling affected such characteristics, since it was clear that data retention and thus refresh was a major technology scaling challenge in DRAM, as indicated by prior and later works (e.g.,~\cite{mandelman.ibmjrd02, kang.memoryforum14, sk-hynix-isscc2023}).

\section{Contributions and Impact}
\vspace{-4pt}

Our paper is the first to comprehensively examine data retention time behavior of modern DRAM chips, uncovering real data and insights on two major phenomena that make retention time identification extremely challenging. Prior works were limited to simulation or had very small sample sizes, and almost none of them examined modern DDR3 DRAM chips or technology scaling. Many device- or circuit-level works did not study DPD or VRT. No architecture- and system-level work to reduce refresh overhead discussed DPD or VRT. Our work enabled a new understanding and demonstrated the true difficulty of a major problem in DRAM technology scaling, by providing valuable data that was available nowhere else (at least publicly).

Our key results demonstrate that data retention times of modern DRAM chips are indisputably getting worse in newer-generation DRAM chips, indicating that refresh is becoming a larger problem with technology scaling. Ditto for DPD and VRT. For example, we showed that 1) the retention failure coverage of a given data pattern becomes smaller for newer-generation DRAM chips, 2) VRT is a widespread phenomenon in modern DRAM devices, causing significant dynamic changes in minimum retention time. These were the first results of their kind. 

Our results indicated that many prior proposals (e.g.,~\cite{asr-refresh, liu2012raidr, ohsawa1998optimizing, multiperiod, venkatesan06, yanagisawa1988semiconductor}) that rely on accurate retention time identification to eliminate refreshes would not work reliably as they do not take into account DPD or VRT. They also put into question whether existing refresh rates are enough to guarantee error-free operation in DRAM chips being used in the field (especially in the presence of VRT). As DRAM technology scales, would it be easy to accurately determine retention times to ensure data integrity even if we maintained a conservative refresh rate for all DRAM cells?  

Based on the understanding we developed, we proposed ideas and avenues for future work on how to tackle the DPD and VRT problems (\$5.2 \& \$6.3 in~\cite{dram-isca2013}). We advocated the use of ECC in DRAM chips to detect and/or correct any retention errors that might not be identified after rigorous testing (offline or online). Most modern DRAM chips now incorporate ECC (see~\cite{patel2019understanding, patel2020beer, patel2021harp}), especially to account for VRT effects~\cite{kang.memoryforum14}. We also advocated the use of online profiling together with ECC to enable reliable identification of retention times, an approach later works rigorously investigated and enabled (e.g.,~\cite{khan.sigmetrics14, avatar-dsn15, khan.dsn16, khan.cal16, khan.micro17, patel2017reaper, patel2020beer, patel2021harp}). As such, our ISCA 2013 paper enabled system-level techniques to overcome a major DRAM scaling challenge, an approach we call {\em system-DRAM co-design}~\cite{mutlu.imw13,mutlu.superfri15}. We believe developing such system-level techniques that can detect and exploit DRAM characteristics online, during system operation, will be increasingly valuable as such characteristics will become much more difficult to accurately determine and exploit due to technology scaling.

A key contribution of our work was the development of our flexible FPGA-based DRAM testing infrastructure, which was the first of its kind. It enabled a large amount of research into DRAM chips by enabling rigorous experimental study of real DRAM chip characteristics, including the rigorous study of the RowHammer vulnerability~\cite{kim_flipping_2014, kim2020revisiting, frigo2020trr, hassan2021utrr, yaglikci2022hira, yaglikci2022understanding, olgun2023drambender, olgun2023hbm, luo2023rowpress, farmani2021rhat, mutlu2020retrospective, mutlu2023fundamentally}, another major DRAM technology scaling problem. We discuss some new insights and studies enabled by this infrastructure in our RAIDR retrospective and our SoftMC~\cite{hassan2017softmc} and DRAM Bender~\cite{olgun2023drambender} works.

\section{Influence on Later Works}
\vspace{-4pt}

Many later works (e.g., ~\cite{khan.sigmetrics14, avatar-dsn15, khan.dsn16, khan.cal16, khan.micro17, patel2017reaper, patel2020beer, patel2021harp}) ensued to solve the DPD \& VRT problems. Some provided a more detailed characterization of the DPD and VRT phenomena:~\cite{khan.sigmetrics14} analyzed both DPD \& VRT and examined the effectiveness of online profiling versus ECC of varying strength. AVATAR~\cite{avatar-dsn15} provided heterogeneous refresh rates using a combination of online profiling, ECC, and memory scrubbing, working from the empirical observation that new VRT errors are discovered infrequently at a steady rate. PARBOR~\cite{khan.dsn16} introduced detailed DPD analyses and a new technique to efficiently detect data-dependent failures. REAPER~\cite{patel2017reaper} analyzed the DPD \& VRT phenomena in newer LPDDR4 DRAM chips, demonstrating that the problems are getting worse, and developed the reach profiling technique to tolerate the two problems. We believe AVATAR \& REAPER enabled practical ways of exploiting heterogeneous retention times.

ECC is mainstream in DRAM chips today~\cite{patel2019understanding, patel2020beer, patel2021harp}. We believe this is a direct result of the analysis that showed the prevalence and importance of VRT and the difficulty of handling VRT-caused retention errors due to their fundamentally unpredictable nature. A later work by Samsung \& Intel engineers~\cite{kang.memoryforum14} described that ECC is needed to deal with VRT, just as our ISCA'13 paper advocated.

\section{Summary and Future Outlook}
\vspace{-4pt}

Our ISCA 2013 paper was a nice example of harmonious collaboration between academia and industry: Intel helped us build the infrastructure and both Intel \& Samsung gave us significant technical feedback along with generous funding. Our paper also highlights the importance of investing into building infrastructure to analyze real chips: doing so enabled not only  the new understanding developed in our work, but also many future works that analyzed various other DRAM characteristics (e.g.,~\cite{lee.hpca15, chang.sigmetrics16, lee2017design, kim2018solar, chang.sigmetrics17,ghose2018vampire, kim_flipping_2014, kim2020revisiting, frigo2020trr, hassan2021utrr, yaglikci2022hira, yaglikci2022understanding, olgun2023drambender, olgun2023hbm, luo2023rowpress, farmani2021rhat, mutlu2020retrospective, mutlu2023fundamentally, khan.sigmetrics14, avatar-dsn15, khan.dsn16, khan.cal16, khan.micro17, patel2017reaper, patel2020beer, patel2021harp}) and uncovered fascinating undocumented capabilities in real DRAM chips, e.g., the ability to perform data copy/initialization and bitwise operations~\cite{Seshadri:2015:ANDOR, seshadri_ambit_2017, pidram, gao2019computedram, gao2022frac}, implement physical unclonable functions~\cite{kim2018dram}, and generate true random numbers~\cite{kim.hpca19,olgun2021quactrng}.

Since ISCA 2013, we have come a long way in understanding fundamental characteristics of DRAM devices and enabling practical solutions to overcome DRAM shortcomings.   Yet, there is a lot more to be empirically discovered and understood in DRAM to solve the fundamental scaling, performance, and energy challenges of the technology (as shown by very recent works in 2022-2023, e.g.~\cite{olgun2023drambender, yaglikci2022hira, yaglikci2022understanding, pidram, gao2022frac, olgun2023hbm, luo2023rowpress}), which can enable solutions also applicable to other technologies. We conclude that the future is bright in experimental memory systems research using real memory chips.

\setstretch{0.70}

\balance
\bibliographystyle{IEEEtran}
{\tiny
\bibliography{combined}}

\end{document}